\documentclass[prl, a4paper, twocolumn, twoside, a4paper,showpacs]{revtex4}
\usepackage{times}
\usepackage{graphicx}
\usepackage{amssymb}
\usepackage{wasysym}
\usepackage{natbib}
\usepackage[usenames,dvipsnames]{color}

\usepackage[bookmarks=true,letterpaper=true,colorlinks=true,urlcolor=blue,linkcolor=blue,citecolor=blue]{hyperref} 
\begin{document}
%\received{11 February 2009}
%\published{22 May 2009}

\title{Computational Design of Chemical Nanosensors: Metal Doped Carbon Nanotubes}
\author{J. M. Garc{\'{\i}}a-Lastra$^{\text{1,2}}$}
\email{juanmaria.garcia@ehu.es}
\author{D. J. Mowbray$^{\text{1,2}}$}
\author{K. S. Thygesen$^{\text{2}}$}
\author{A. Rubio$^{\text{1,3}}$}
\author{K. W. Jacobsen$^{\text{2}}$}
\affiliation{
$^{\text{1}}$Nano-Bio Spectroscopy group and ETSF Scientific Development Centre, Dpto. F{\'{\i}}sica de Materiales, Universidad del Pa{\'{\i}}s Vasco, Centro de F{\'{\i}}sica de Materiales CSIC-UPV/EHU- MPC and DIPC, Av. Tolosa 72, E-20018 San Sebasti{\'{a}}n, Spain\\
$^{\text{2}}$Center for Atomic-scale Materials Design, Department of Physics,
  Technical University of Denmark, DK-2800 Kgs.~Lyngby, Denmark\\
$^{\text{3}}$Fritz-Haber-Institut der Max-Planck-Gesellschaft, Berlin, Germany
} 

\begin{abstract}
  We use computational screening to systematically investigate the use
  of transition metal doped carbon nanotubes for chemical gas sensing.
  For a set of relevant target molecules (CO, NH$_{\text{3}}$,
  H$_{\text{2}}$S) and the main components of air
  (N$_{\text{2}}$, O$_{\text{2}}$, H$_{\text{2}}$O), we calculate the
  binding energy and change in conductance upon adsorption
  on a metal atom occupying a vacancy of a (6,6) carbon nanotube. Based on
  these descriptors, we identify the most promising dopant candidates
  for detection of a given target molecule. From the fractional
  coverage of the metal sites in thermal equilibrium with air, we
  estimate the change in the nanotube resistance per doping site as a
  function of the target molecule concentration assuming charge
  transport in the diffusive regime. Our analysis points to Ni-doped
  nanotubes as candidates for CO sensors working under  typical
  atmospheric conditions.  
\end{abstract}

\maketitle

The ability to detect small concentrations of specific chemical
species is fundamental for a variety of industrial and scientific
processes as well as for medical applications and environmental
monitoring \cite{gas_sensing}. In general, nanostructured materials
should be well suited for sensor applications because of their large
surface to volume ratio which makes them sensitive to molecular
adsorption. Specifically, carbon nanotubes (CNT) \cite{cnt_review} have
been shown to work remarkably well as detectors of small gas
molecules. This has been demonstrated both for individual
CNTs \cite{kong,collins,hierold,villalpando,rocha,Brahim} as well as
for CNT networks \cite{morgan,cnt_networks}. 

Pristine CNTs are known to be chemically inert -- a property closely
related to their high stability. As a consequence, only radicals bind
strong enough to the CNT to notably affect its electrical properties
\cite{cnt_review,hierold,Valentini2004356,zanolli:155447,Juanma}. To
make CNTs attractive for sensor applications thus requires some kind
of functionalization, e.g. through doping or decoration of the CNT
sidewall \cite{Fagan, Yagi, Yang, Chan, Yeung, Vo, Furst,Juanma,
  Krasheninnikov}. Ideally, this type of functionalization could be
used to control not only the reactivity of the CNT but also the
selectivity towards specific chemical species. 

In this work we consider the possibility of using CNTs doped by 3d
transition metal atoms for chemical gas sensing. We use computational
screening to systematically identify the most promising dopant
candidates for detection of three different target molecules (CO,
NH$_{\text{3}}$, H$_{\text{2}}$S) under typical atmospheric
conditions. The screening procedure 
is based on the calculation of two microscopic descriptors: the
binding energy and scattering resistance of the molecules when
adsorbed on a doped CNT.  These two quantities give a good indication of
the gas coverage and impact on the resistance. For the
most promising candidates we then employ a simple thermodynamic model
of the CNT sensor. In this model, the binding energies are used to obtain
the fractional coverage of the metallic sites as a function of the
target molecule concentration under ambient conditions. Under the assumption of
transport in the diffusive rather than localization regime, the
change in CNT resistivity may then be obtained from the calculated coverages
and single impurity conductances.

We find that oxidation of the active metal site passivates the sensor
in the case of doping by Ti, V, Cr, and Mn under standard conditions
(room temperature and 1 bar of pressure). Among the remaining metals,
we identify Ni as is the most promising candidate for CO detection.
For this system the change in resistance per active site is generally
significant ($>$1~$\Omega$) for small changes in CO concentration in
the relevant range of around 0.1--10~ppm. Our approach is quite
general and is directly applicable to other nanostructures than CNTs,
other functionalizations than metal doping, and other backgrounds than
atmospheric air.

All total energy calculations and structure optimizations have been
performed with the real-space density functional theory (DFT) code
\textsc{gpaw}  \cite{GPAW} which is based on the projector augmented
wave method. We use a grid spacing of 0.2~\AA~for representing the
density and wave functions and the PBE exchange correlation functional
\cite{PBE}.  Transport calculations for the optimized structures have
been performed using the non-equilibrium Green's function method
\cite{benchmark} with an electronic Hamiltonian obtained from the
\textsc{siesta} code \cite{SIESTA} in a double zeta polarized (DZP)
basis set. Spin polarization has been taken into account in all
calculations. 

\begin{figure}[t]
\includegraphics[width=0.975\columnwidth]{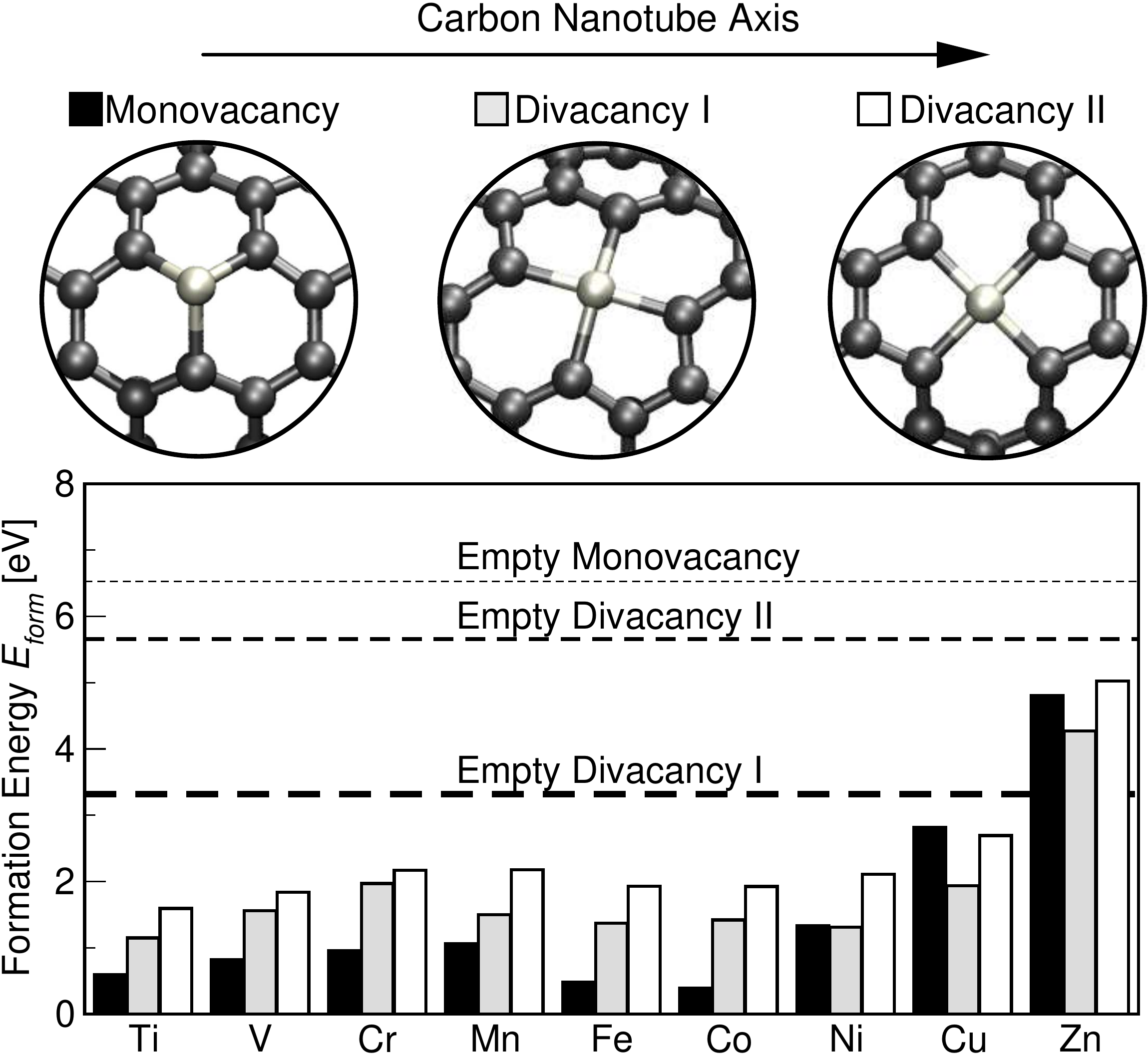}
\caption{Structural schematics and formation energy for a 3d
  transition metal occupied monovacancy (black), divacancy I (gray),
  or divacancy II (white) in a (6,6) carbon nanotube. Formation
  energies of the empty vacancies are indicated by dashed
  lines.}\label{Fig1}  
\end{figure}

Metallic doping of a (6,6) CNT has been modeled in a supercell
containing six repeated minimal unit cells along the CNT axis
(dimensions: 15 \AA$\times$15 \AA$\times$14.622 \AA). For this size of
supercell a \(\Gamma\)-point sampling of the Brillouin zone was found
to be sufficient.  The formation energy for creating a vacancy (VC)
occupied by a transition metal atom (M) was calculated using the
relation 
\begin{equation}
E_{\text{form}}[\text{M}@\text{VC}] = E[\text{M}@\text{VC}] +
n E[\text{C}] - E[\text{M@NT}] 
\end{equation}
where $E$[M@VC] is the total energy of a transition metal atom
occupying a vacancy in the nanotube, $n$ is the number of carbon atoms
removed to form the vacancy, $E$[C] is the energy per carbon atom in
a pristine nanotube, and $E$[M@NT] is the total energy of the pristine
nanotube with a physisorbed transition metal atom. We have considered
the monovacancy and two divacancies shown in Fig.~\ref{Fig1}.  The
energy required to form an empty vacancy is obtained from 
\begin{equation}
E_{\text{form}}[\text{VC}] = E[\text{VC}] + n
E[\text{C}] - E[\text{NT}], 
\end{equation}
where $E$[VC] is the total energy of the nanotube with a vacancy of
$n$ atoms. 

The calculated formation energies for the 3d transition metals are
shown in Fig.~\ref{Fig1}. From the horizontal lines we see that both
divacancies are more stable than the monovacancy. This may be
attributed to the presence of a two-fold coordinated C atom in the
monovacancy, while all C atoms remain three-fold coordinated in the
divacancies.  When a transition metal atom occupies a vacancy, the
strongest bonding to the C atoms is through its $d$
orbitals~\cite{Griffith}. For this reason, Cu and Zn, which both have
filled d-bands, are rather unstable in the CNT. For the remaining
metals, adsorption in the monovacancies leads to quite stable
structures. This is because the three-fold coordination of the C atoms
and the CNT's hexagonal structure are recovered when the metal atom is
inserted. On the other hand, metal adsorption in divacancies is
slightly less stable because of the resulting pentagon defects, see
upper panel in Fig.~\ref{Fig1}.  A similar behaviour has been reported
by Krasheninnikov \emph{et al.} for transition metal atoms in graphene
\cite{Krasheninnikov}. 

\begin{figure}[t]
\includegraphics[width=\columnwidth]{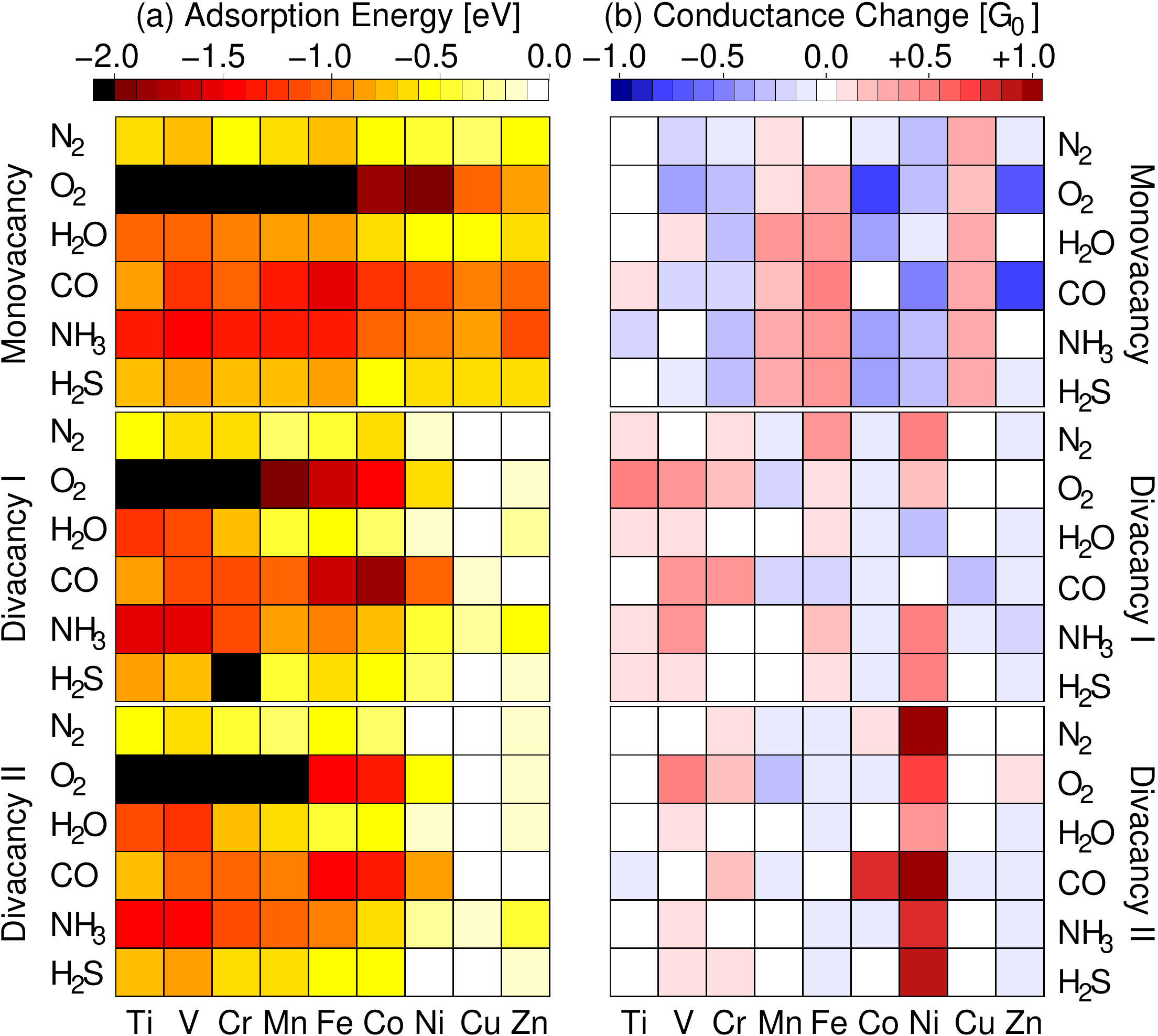}
\caption{Calculated (a) adsorption energy $E_{\text{ads}}$ in eV and
  (b) change in conductance $\Delta G$ in units of $G_0 =
  $2$e^{\text{2}}/h$ for N$_{\text{2}}$, O$_{\text{2}}$,
  H$_{\text{2}}$O, CO, NH$_{\text{3}}$, and H$_{\text{2}}$S on 3d
  transition metals occupying a monovacancy (top), divacancy I
  (middle), and divacancy II (bottom) in a (6,6) carbon
  nanotube.}\label{Fig2}  
\end{figure}

The adsorption energies for N$_{\text{2}}$, O$_{\text{2}}$,
H$_{\text{2}}$O, CO, NH$_{\text{3}}$, and H$_{\text{2}}$S on the
metallic site of the doped (6,6) CNTs are shown in 
Fig.~\ref{Fig2}(a).  The adsorption energy of a molecule $X$ is defined by 
\begin{equation}
E_{\text{ads}}[X\text{@M@VC}] = E[X\text{@M@VC}] - E[X] - E[\text{M@VC}], 
\end{equation}
where $E$[$X$@M@VC] is the total energy of molecule $X$ on a
transition metal atom occupying a vacancy, and $E[X]$ is the gas phase
energy of the molecule. 

From the adsorption energies plotted in Fig.~\ref{Fig2}(a), we see that
the earlier transition metals tend to bind the adsorbates stronger
than the late transition metals. The latest metals in the series (Cu
and Zn) bind adsorbates rather weakly in the divacancy structures. We 
also note that O$_{\text{2}}$ binds significantly stronger than any of the
three target molecules on Ti, V, Cr, and Mn (except for Cr in
divacancy I where H$_{\text{2}}$S is found to dissociate).  Active sites
containing these metals are therefore expected to be completely
passivated if oxygen is present in the background. Further, we find
H$_{\text{2}}$O is rather weakly bound to most of the active sites.
This ensures that these types of sensors are robust against changes in
humidity. 

In thermodynamic equilibrium \cite{textbook}, the coverage of the
active sites follows from   
\begin{eqnarray}
\Theta[X] & = & \frac{K[X] C[X]}{1 + \sum_{Y} K[Y] C[Y]},
\end{eqnarray}
where $K = k_+ / k_-$ is the ratio of forward and backward
rate constants for the adsorption reaction,
\begin{eqnarray}
K[X] & = & \exp \left[-\frac{ E_{\text{ads}}[X] + T S[X]}{k_B T}\right].
\end{eqnarray}
In these expressions $C[X]$ is the concentration of species $X$,
$S[X]$ is its gas phase entropy and $T$ is the
temperature. Experimental values for the gas phase entropies have been
taken from Ref.~\cite{CRCHandbook}. 

\begin{figure}[t]
\includegraphics[width=\columnwidth]{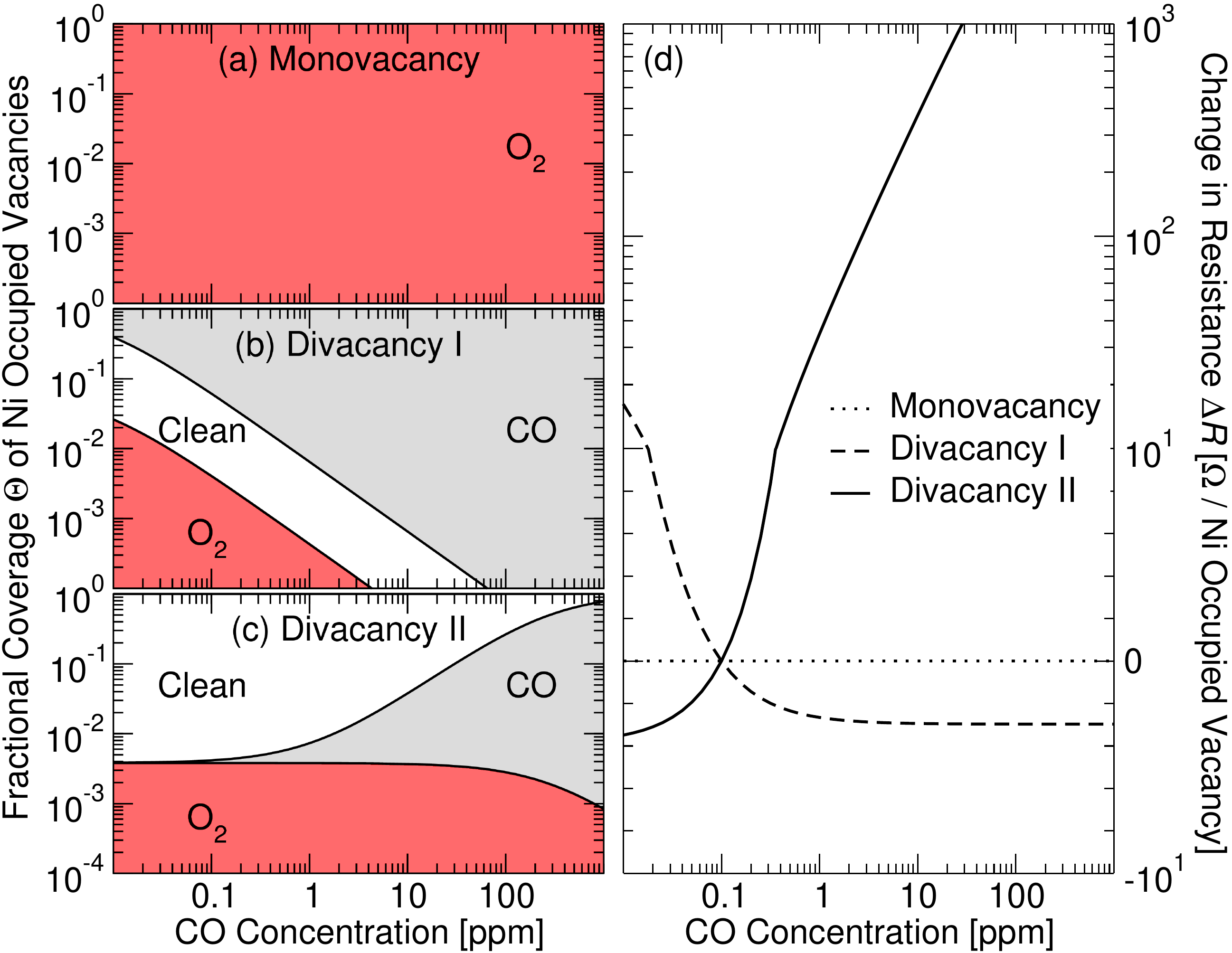}
\caption{Fractional coverage \(\Theta\) in thermal equilibrium of Ni
  in a (a) monovacancy, (b) divacancy I, (c) divacancy II and (d)
  change in resistance \(\Delta R\) per dopant site as a function of
  CO concentration in a background of air at room temperature and 1
  bar of pressure. The reference concentration of CO is taken to be
  $C_0=$0.1 ppm. Note the change from linear to log scale on the
  $y$-axis at $\Delta R=$10~$\Omega$.}\label{Fig3}  
\end{figure}

For a given background composition we may thus estimate the fractional
coverages for each available adsorbate for a given type of doping. As
an example, Fig.~\ref{Fig3}(a)-(c) shows the fractional coverage of a Ni atom
occupying a monovacancy, divacancy I, and divacancy II, versus CO
concentration in a background of air at room temperature and 1 bar of
pressure.  Due to the relatively small binding energy of N$_{\text{2}}$ and
H$_{\text{2}}$O as compared to O$_{\text{2}}$ and CO, all Ni sites
will be either empty or occupied by O$_{\text{2}}$ or CO. In
particular, Ni in a monovacancy (top panel of Fig.~\ref{Fig3}) will be
completely oxidized for all relevant CO concentrations.  For the Ni
occupied divacancy II structures we find the coverage of CO changes
significantly around toxic concentrations ($\sim$10 ppm).

To estimate the effect of adsorbates on the electrical conductance of
doped CNTs, we first consider the change in conductance when a
single molecule is adsorbed on a metal site of an otherwise pristine
CNT.  In Fig.~\ref{Fig2}(b) we show the calculated change in conductance
relative to the metal site with no adsorbate. In contrast to the
binding energies, there are no clear trends in the conductances. The
sensitivity of the conductance is perhaps most clearly demonstrated by the
absence of correlation between different types of vacancies,
i.e.~between the three panels in Fig.~\ref{Fig2}(b). Close to 
the Fermi level, the conductance of a perfect armchair CNT equals
2$G_0$.  The presence of the metal dopant leads to several dips in the
transmission function known as Fano antiresonances  \cite{Furst}. The
position and shape of these dips depend on the $d$-levels of the
transition metal atom, the character of its bonding to the CNT, and
is further affected by the presence of the adsorbate molecule. The coupling of
all these factors is very complex and makes it difficult to estimate
or rationalize the value of the conductance. For the spin polarized
cases, we use the spin-averaged conductances,
i.e.~$G=(G_{\uparrow}+G_{\downarrow})/{\text{2}}$. 

Next, we estimate the resistance of a CNT containing several impurities
(a specific metal dopant with different molecular adsorbates).  Under
the assumption that the electron phase-coherence length, $l_\phi$, is
smaller than the average distance between the dopants, $d$, we may
neglect quantum interference and obtain the total resistance by adding
the scattering resistances due to each impurity separately. The
scattering resistance due to a single impurity is given by
\begin{eqnarray}
R_s(X)&=&{\text{1}}/G(X)-{\text{1}}/({\text{2}}G_0),
\end{eqnarray}
where $G(X)$ is the Landauer conductance of the pristine CNT with a
single metal dopant occupied by molecule $X$ and
${\text{1}}/({\text{2}}G_0)$ is the contact resistance of a (6,6)
CNT.

We may now obtain the total resistance per dopant site relative to the
reference background signal as a function of the target molecule
concentration 
\begin{eqnarray}\label{eq.totalres}
\frac{\Delta R}{N} &\approx& \sum_{X} R_s(X)(\Theta[X,C] - \Theta[X,C_0]), 
\end{eqnarray}
where $N$ is the number of dopants, $\Theta[X,C]$ is the
fractional coverage of species $X$ at concentration $C$ of the
target and $C_0$ is the reference concentration. Notice
that the contact resistance drops out as we evaluate a change in
resistance.

In Fig.~\ref{Fig3}(d) we show the change in resistance calculated from
Eq. (\ref{eq.totalres}) as a function of CO concentration for Ni
occupying the three types of vacancies.  The background reference
concentration of CO is taken to be $C_0=0.1$ ppm. For the monovacancy
there is very little change in resistivity. This is because most
active sites are blocked by O$_{\text{2}}$ at relevant CO concentrations, as
shown in the upper panel of Fig.~\ref{Fig3}. For Ni in the divacancies
there is, however, a change in resistance on the order of $1\Omega$
per site. For concentrations above \(\sim\)1 ppm, the CO coverage of
Ni in the divacancy II increases dramatically and this leads to a
significant increase in resistance.  

We now return to the discussion of the validity of
Eq.~(\ref{eq.totalres}). As mentioned, the series coupling of
individual scatterers should be valid when $l_\phi<d$. However, even
for $l_\phi>d$ and assuming that the Anderson localization length,
$l_{\text{loc}}$ in the system exceeds $l_\phi$,
Eq.~(\ref{eq.totalres}) remains valid if one replaces the actual
resistance $R$ by the sample averaged resistance $\langle R  
\rangle$ \cite{Brandbyge}. At room temperature under ambient
conditions, interactions with external degrees of freedom such as
internal CNT phonons and vibrational modes of the adsorbed molecules
would rapidly randomize the phase of the electrons. Therefore Eq.
(\ref{eq.totalres}) should certainly be valid in the limit of low
doping concentrations. On the other hand, the total number of dopants,
$N$, should be large enough for the statistical treatment of the
coverage to hold. Finally, we stress that Eq. (\ref{eq.totalres})
represents a conservative estimate of the change in resistance. In
fact, in the regime where $l_\phi>l_{\text{loc}}$, i.e.~in the
Anderson localization regime, the resistance would be highly sensitive
to changes in the fractional coverage of active sites. Calculation of
the actual resistance of the CNT in this regime would, however,
involve a full transport calculation in the presence of all $N$
impurities. At this point it suffices to see that the conservative
estimates obtained from Eq.~(\ref{eq.totalres}) predict measurable
signals in response to small changes in concentration of the target molecules.

To our knowledge, controlled doping of CNTs with transition metal
atoms has so far not been achieved.  It has, however, been found that
metal atoms incorporated into the CNT lattice during catalytic growth
are afterwards very difficult to remove~\cite{Ushiro}. Furthermore, it
has been shown that CNT vacancies, which are needed for the metallic
doping, may be formed in a controlled way by irradiation by Ar
ions \cite{navarro}. This suggests that metallic doping of CNTs should
be possible. 

In summary, we have presented a general model of nanostructured
chemical sensors which takes the adsorption energies of the relevant
chemical species and their individual scattering resistances as the
only input. On the basis of this model we have performed a
computational screening of transition metal doped CNTs, and found that
Ni-doped CNTs are promising candidates for detecting CO in a background of air. The model may be
applied straightforwardly to other nanostructures than CNTs, other
functionalizations than metal doping and other gas compositions than air.

\acknowledgments

The authors acknowledge financial support from 
Spanish MEC (FIS2007-65702-C02-01),
``Grupos Consolidados UPV/EHU del Gobierno Vasco'' (IT-319-07), e-I3 ETSF
project (Contract Number 211956), ``Red Espa{\~{n}}ola de
Supercomputaci{\'{o}}n'', NABIIT and the Danish Center for Scientific
Computing. The Center for Atomic-scale Materials Design (CAMD) is
sponsored by the Lundbeck Foundation.  JMG-L acknowledges funding from
Spanish MICINN through Juan de la Cierva and Jos{\'{e}} Castillejo
programs. 

%\bibliographystyle{arXiv}
%\bibliography{../PRL/bibliography}

\end{document}